\documentclass[11pt,twoside]{article}
\usepackage{./asp2014}

\aspSuppressVolSlug
\resetcounters

\begin{document}

\title{MOS: A critical tool for current and future radio surveys}
\author{Daniel J.B. Smith
\affil{Centre for Astrophysics Research, University of Hertfordshire, College Lane, Hatfield, Herts, AL10 9AB, United Kingdom; \email{daniel.j.b.smith@gmail.com}}}

% This section is for ADS Processing.  There must be one line per author.
\paperauthor{Daniel~J.B.~Smith}{daniel.j.b.smith@gmail.com}{}{University of Hertfordshire}{Centre for Astrophysics Research}{Hatfield}{Herts}{AL10 9AB}{United Kingdom}

\begin{abstract}
Since radio continuum observations are not affected by dust obscuration, they are of immense potential diagnostic power as cosmological probes and for studying galaxy formation and evolution out to high redshifts. However, the power-law nature of radio frequency spectra ensures that ancillary spectroscopic information remains critical for studying the properties of the faint radio sources being detected in rapidly-increasing numbers on the pathway to the Square Kilometre Array. In this contribution, I present some of the key scientific motivations for exploiting the immense synergies between radio continuum observations and multi-object spectroscopic surveys. I review some of the ongoing efforts to obtain the spectra necessary to harness the huge numbers of star-forming galaxies and AGN that current and future radio surveys will detect. I also touch on the WEAVE-LOFAR survey, which will use the WEAVE spectrograph currently being built for the William Herschel Telescope to target hundreds of thousands of low-frequency sources selected from the LOFAR continuum surveys.
\end{abstract}

\section{Introduction}
\label{sec:intro}

Radio surveys are rapidly becoming more sensitive over increasingly large areas in the build up to the construction phase of the Square Kilometre Array (hereafter the SKA). The increased capabilities that the SKA will bring are so far ahead of existing facilities that several pathfinders and precursor telescopes are being built, including the Australian SKA Pathfinder \citep[ASKAP;][]{norris11}, MeerKAT \citep[][]{jarvis12} and the Low Frequency Array \citep[LOFAR;][]{vanhaarlem13}. In the case of LOFAR, the huge increase in sensitivity is complemented by low frequency coverage, opening up one of the last largely unexplored areas of the electromagnetic spectrum. Since they are not affected by dust obscuration, radio continuum (hereafter `RC') surveys offer exquisite sensitivity to key processes governing galaxy formation and evolution that are often associated with dusty -- and therefore highly obscured -- environments; active galactic nuclei (hereafter `AGN') and star formation (`SF'). These two phenomena are intimately linked, and this is highlighted by the fact that both AGN and SF peaked in activity around $1< z < 2$ and have subsequently declined \citep[e.g.][etc]{madau98,hopkins06,delvecchio14}, as well as by the observed correlations between black hole masses and properties of the host galaxy \citep[e.g.][]{ferrarese00,gebhardt00,haring04}. Indeed, AGN activity is thought to play a critical role in regulating SF in massive galaxies \citep[e.g.][]{silk98,hopkins08}, whether that takes the form of a total truncation (the so-called `quasar', `high-excitation' or `efficient' mode) or a less dramatic `maintenance' (or alternatively `radio', `low-excitation' or `inefficient') mode \citep[see e.g.][for recent discussions]{best12,gurkan14}. 

In what follows, we highlight the immense synergy between new and forthcoming RC surveys and multi-object spectrographs (MOS), and attempt to provide some insight as to why MOS surveys are a critical tool for extracting the cutting-edge science that is becoming possible as a result. 

\articlefigure[width=0.9\textwidth, trim=0.2cm 0.1cm 0.3cm 0cm, clip=true]{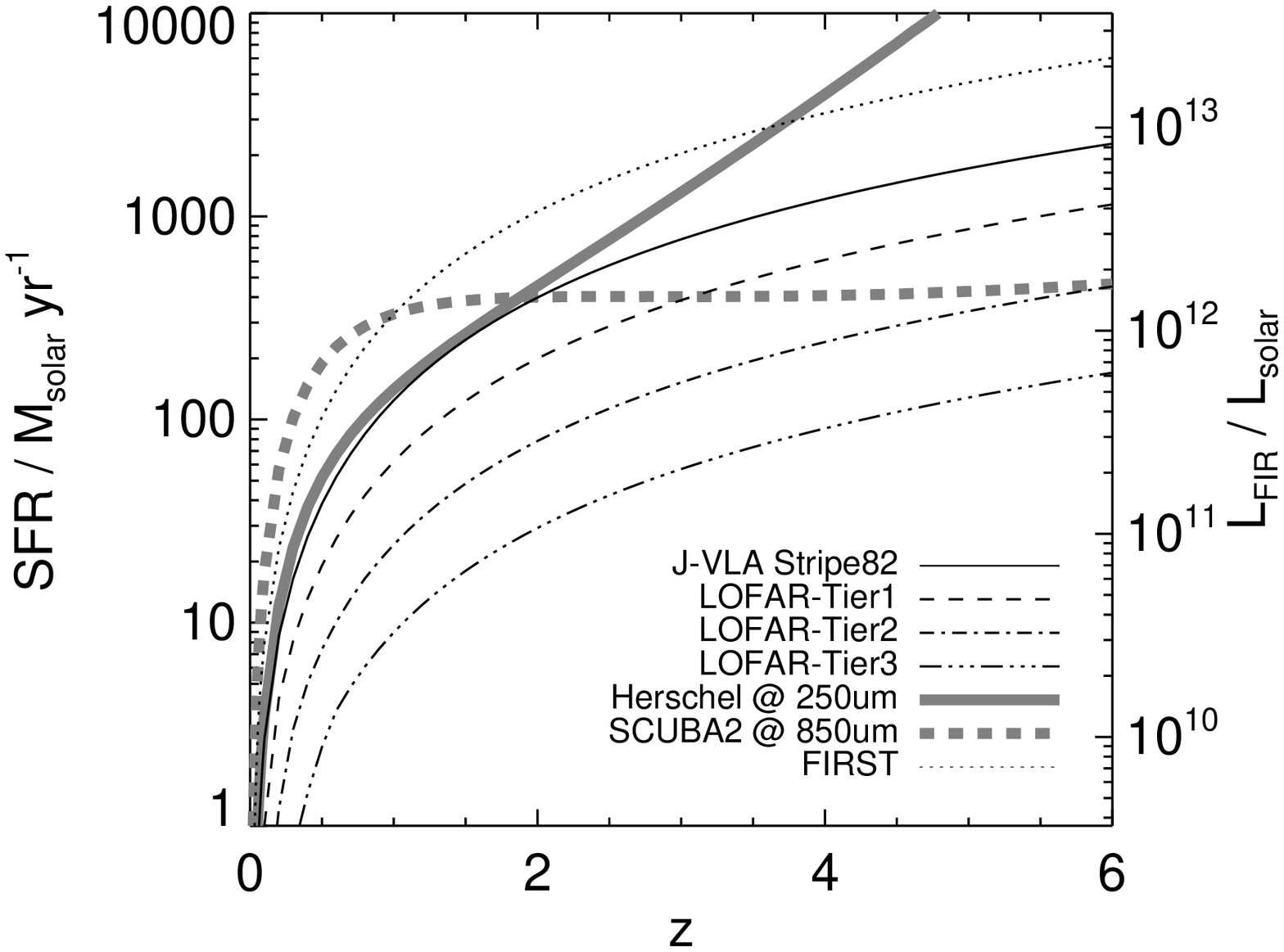}{fig:SFR_comparisons}{Star formation rate sensitivity of selected far-infrared observatories and radio continuum surveys as a function of redshift. The thick grey lines show the confusion-limited SFR sensitivity of the {\it Herschel Space Observatory} at 250\,$\mu$m and the SCUBA-2 instrument on the JCMT at 850\,$\mu$m (solid and dashed gray lines, respectively). Overlaid in black are curves detailing the SFR sensitivity of radio surveys from the J-VLA over Stripe 82 (Heywood et al. {\it in prep}; solid line), and the three tiers of the LOFAR Surveys KSP; Tier 1 (dotted line), Tier 2 (dot-dashed line), and Tier 3 (dot-dot-dot-dashed).}

\section{The power of RC for studying active galaxies}

\noindent Figure \ref{fig:SFR_comparisons} shows the relative star formation rate sensitivity of selected SKA pathfinder surveys as a function of redshift, including the three tiers of the LOFAR Surveys Key Science Project \citep{rottgering11} and the J-VLA survey of the Sloan Digital Sky Survey Stripe 82 region from Heywood et al. ({\it in prep}). Tier 1 of the  LOFAR surveys (dashed line in figure \ref{fig:SFR_comparisons}) will cover the whole northern sky with SFR sensitivity greater than the $5\sigma$ confusion limit of {\it Herschel} \citep[][shown as the solid gray line]{pilbratt10}. In addition, the extremely faint confusion limits, coupled with the dramatic sensitivity afforded by long integration times, ensures that the ultra-deep Tier 3 will cover tens of square degrees to depths even greater than the most sensitive existing RC imaging \citep[e.g.][]{schinnerer07,schinnerer10,smolcic14}, far outstripping the confusion-limited SFR sensitivity of SCUBA-2 \citep[][dashed thick gray line]{holland13}.\footnote{These SFRs have been calculated by taking the surveys' respective $5\sigma$ flux density limits and using the radio-luminosity to star formation rate calibration from \citet{bell03} from $0 < z < 6$ and assuming a spectral index, $\alpha = 0.7$ (we use the convention that $S_\nu \propto \nu^{-\alpha}$) following \citet{mauch13}. The far-IR SFR values in figure \ref{fig:SFR_comparisons} assume a simple isothermal dust model \citep[e.g.][]{hildebrand83,hayward12,smith13}; they are likely to over-estimate the star formation rate sensitivity of {\it Herschel} and SCUBA-2.} It is clear from figure \ref{fig:SFR_comparisons} that the sensitivity of Tier 3 of the LOFAR surveys is such that it will allow the routine detection of SMG-like galaxies at $z > 5$.

Of course, RC surveys are also optimal for studying sources containing AGN. Indeed, radio frequencies are the ideal way to select complete samples of AGN out to high redshift due to their sensitivity to the highly obscured type-II sources that are missed by optical\slash near-IR studies \citep[e.g.][]{maddox12}. This sensitivity includes access to the X-ray invisible Compton-thick sources, and it has been suggested that they could even outnumber the unobscured population at high redshifts \citep[e.g.][]{alejo07}. However, to exploit the power of the SKA pathfinders, it is essential that: (i) redshifts can be estimated, and (ii) the star-forming sources can be separated from those whose RC emission is dominated by AGN. These two aspects are discussed in the coming sections. 

\section{The need for spectra and the nature of the faint radio source population}

Redshift information is critical if we are to harness the immense sensitivity of the SKA pathfinders (detailed in figure \ref{fig:SFR_comparisons}) to study the co-evolution of the intimately-linked processes of AGN activity and SF. Unfortunately, though the new generation radio data sets often include multiple frequencies, the power-law nature of radio source spectra (whether they are dominated by synchrotron or free-free emission, associated with star formation or AGN) ensures that the RC data sets alone do not contain any redshift information. It must be derived by other means. 

For many tasks, such as simply studying the cosmic star formation rate density evolution, redshifts derived from optical to near-infrared photometry may be sufficient \citep[photometric redshifts are discussed in detail in e.g.][]{bolzonella00,collister04,ilbert06,smith11,almosallam15}. However, photometric redshifts are unreliable for individual sources, particularly for those which are dusty, have bright emission lines, or contain AGN. These groups comprise a large component of sources observed at faint RC flux densites, which are selected based on their {\em activity}; as a result, it is only by obtaining spectra for large samples of RC-selected that we can harness the immense diagnostic power of RC surveys. The bright emission lines that preclude the estimation of reliable photometric redshifts are extremely useful when it comes to obtaining spectroscopic redshifts for radio sources; their presence means that redshifts can be obtained from relatively short spectroscopic exposures since it is not necessary to detect continuum emission. This fact, combined with the massive multiplex of forthcoming MOS instruments -- including e.g. DESI \citep{flaugher14}, 4MOST \citep[e.g.][]{dejong11}, MOONS \citep[e.g.][]{cirasuolo11}, the MSE \citep{mcconnachie14} and WEAVE \citep[][see below]{dalton12} -- means that MOS follow-up of radio sources is a highly efficient way to obtain redshifts.

\articlefigure[width=0.9\textwidth]{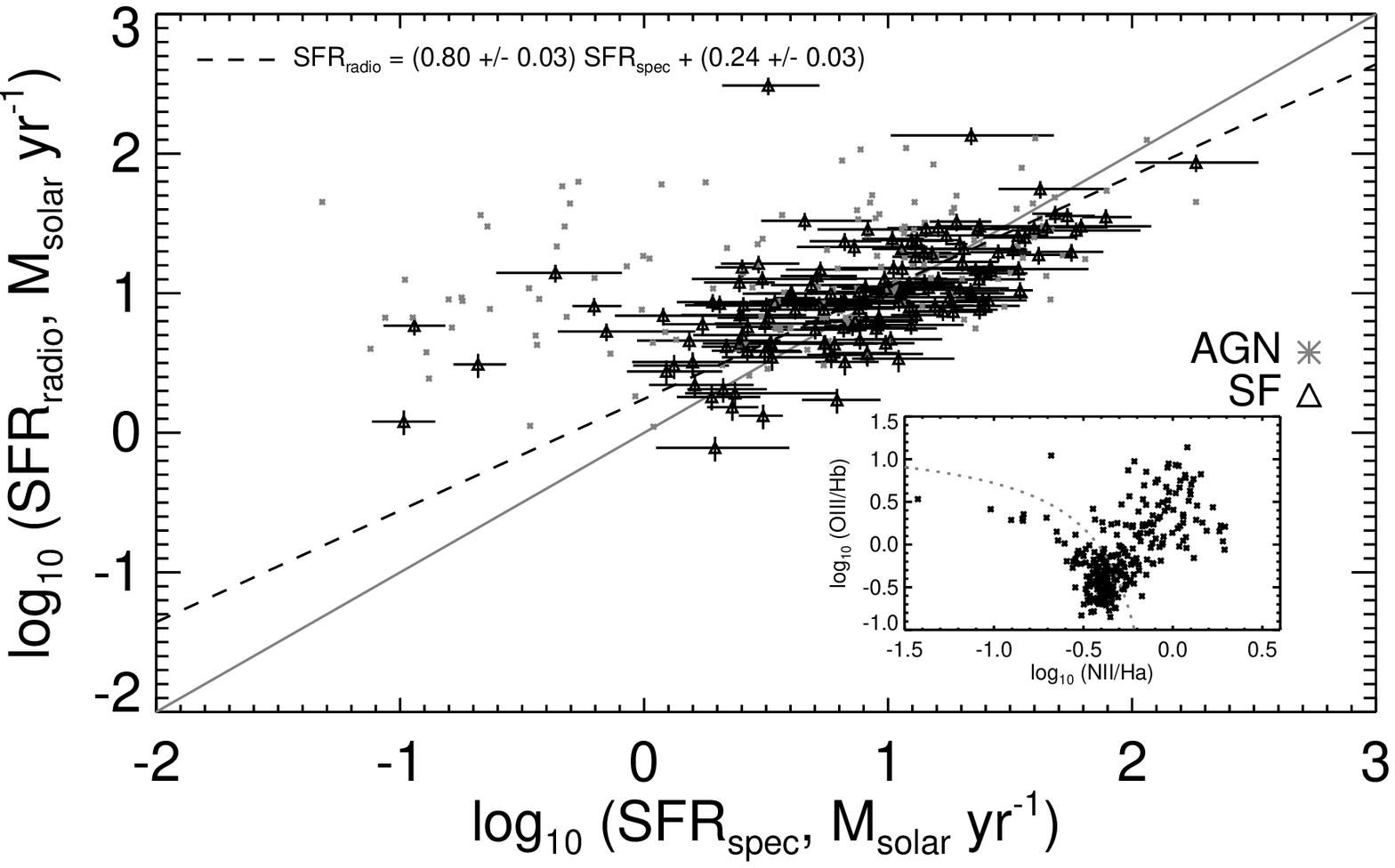}{fig:SFR_bpt}{Comparison between SFR indicators derived for radio sources based on the \citet{bell03} radio-luminosity--SFR calibration and the MPA-JHU Sloan Digital Sky Survey value-added spectroscopic catalogue from \citet{brinchmann04}, and separating SF and AGN sources using the BPT diagram \citep[inset;][]{bpt81}.}

The faint radio source population is known to be diverse \citep[e.g.][]{simpson06,condon12}, and to be comprised of star forming galaxies (SFGs) as well as radio-loud and radio-quiet AGN (RL-AGN and RQ-AGN). Determining the relative mix of star forming and AGN-powered sources at the faint flux densities that the new RC data will explore is critical for studying the interplay between SF and accretion. Though some AGN can be identified using RC morphological information alone, large multiplex MOS spectrographs are critical to be able to do this for the large majority of sources which are unresolved. Spectroscopy provides a means of cleanly separating the star forming sources from the AGN population using emission line diagnostics \citep[e.g.][]{bpt81}, as well as a means to separate different accretion modes \citep[i.e. HERGs and LERGs, see e.g.][for details]{laing94,mauch07,hardcastle13}. Figure \ref{fig:SFR_bpt} shows a comparison between SFRs estimated from the MPA-JHU SDSS value-added spectroscopic catalogue \citep{brinchmann04}, and those estimated using the 1.4\,GHz J-VLA Stripe 82 catalogue from Heywood et al. ({\it in prep}), along with the SDSS spectroscopic redshifts and the \citet{bell03} SFR calibration of radio luminosity. It is clear that those sources classified as AGN using the BPT diagram (inset, and shown as asterisks in the main panel) show an excess 1.4\,GHz luminosity beyond what would be expected from star formation. For those sources classified as star-forming galaxies in the BPT diagram (triangles with error bars overlaid in figure \ref{fig:SFR_bpt}), the best fit relation between the optical and RC SFRs (shown as the dashed line, and parameterised in the top left legend) has a gradient that is in tension with unity (i.e. the SFRs do not agree 1:1, which is shown by the solid line). This may be explained either by e.g. the influence of obscuration on the spectroscopic SFR calibration, or variation in the obscuration-independent RC luminosity--SFR calibration at the faint flux densities probed by the J-VLA data. One possible reason for the latter explanation might be variation in the far-infrared radio correlation at faint RC flux densities \citep[e.g.][McAlpine et al. {\it in prep}]{smith14}.

\section{Current and future MOS surveys of radio sources}

Current spectroscopic efforts targeting RC-selected samples include the SDSS-III BOSS redshift survey of LOFAR sources in the ELAIS N1 field\footnote{\url{http://www.sdss.org/dr12/algorithms/ancillary/boss/sdsslofar/}} (which also targeted sources selected from other sources, including FIRST, GMRT and J-VLA data). In addition, the Spectroscopic Survey of Stripe82 Radio Sources, ``S$^4$", is an international project led from the UK and Netherlands using the AF2+WYFFOS multi-object spectrograph on the William Herschel Telescope (the WHT)\footnote{The S$^4$ website is \url{http://star.herts.ac.uk/~dsmith/Dan\_Smiths\_Website/S4.html}}. S$^4$ (Smith et al. {\it in prep}) aims to obtain 2000 redshifts of radio sources brighter than $\sim 200\,\mu$Jy selected from the 1.4\,GHz VLA and J-VLA data from \citet{hodge11} and Heywood et al. (in prep) over 20\,deg$^2$ within the Stripe 82 region. The first data were taken over four nights during September 2014.

Perhaps the most exciting forthcoming source of spectroscopic follow-up of RC sources is the WEAVE-LOFAR survey (Smith et al. {\it in prep}), which will use the WEAVE instrument \citep{dalton12}. WEAVE\footnote{\url{http://www.ing.iac.es/weave/}} is a MOS instrument being built for the WHT, with 1000 fibres which can be positioned over a field of view 2\,deg in diameter, and continuous wavelength coverage between approximately $0.37 < \lambda (\mu m) < 1.00$ at $R = 5000$. WEAVE first light is scheduled for 2017, and the details of the WEAVE-LOFAR survey are being defined, but we plan to use a `wedding-cake' strategy to target hundreds of thousands of sources selected from the LOFAR surveys, to realise a wide range of science goals. These include: 

\begin{itemize}
\item determining the nature of the faint radio source population at ultra-faint flux limits, 
\item studying the relationship between star formation and accretion over cosmic history, 
\item studying the interplay between different accretion modes, 
\item probing the assembly history of galaxies as a function of mass, environment and type, 
\item probing the influence of environment on SF and AGN activity, 
\item determining the physics of the far-infrared radio correlation,
\item investigating the mechanism behind the RL--RQ AGN dichotomy,
\item providing critical redshift information for supernova cosmology,
\item answering questions about the origin of cosmic magnetism, and
\item finding statistical samples of RL-AGN at $z > 2$ and into the epoch of reionisation at $z > 6$ for 21cm absorption experiments. 
\end{itemize}

\noindent As discussed above, spectra enable much more than simply determining redshifts; it will also be possible to estimate velocity dispersions, metallicities, virial black hole masses, and so on, for a large subset of the RC-selected targets. The spectra that WEAVE-LOFAR will produce will be so numerous and uniform that it will be possible to use stacking techniques to probe the optical continuum properties of galaxies and AGN of those sources for which the continuum signal-to-noise ratio is too low in the individual spectra. This will also enable us, for example, to search for statistical evidence of feedback \citep[e.g.][]{chen10} as a function of type, mass, environment, redshift, and SFR. The spectroscopic data set will also be ideal for conducting statistical reverberation mapping experiments \citep[e.g.][]{fine13}, allowing detailed studies of accretion disk structures. These aspects of the WEAVE-LOFAR survey will be discussed in Smith et al. ({\it in prep}). 

\section{Conclusions}

Though current and future radio continuum surveys are exquisitely sensitive to star formation and AGN activity, and though this sensitivity is undiminished by the presence of dust, the power-law spectral energy distributions of galaxies and AGN at radio frequencies means that they contain no redshift information. If we are to harness the huge diagnostic power that RC surveys possess, redshifts must be derived at other wavelengths, and because of the limitations of photometric redshifts for these sources, MOS represents a prime way to do this. Spectroscopy also enables a whole raft of other measurements, including allowing RC sources to be classified (e.g. as star-forming or AGN, radio-quiet or radio-loud, etc). RC sources frequently have bright emission lines, which (along with the large multiplex of MOS) makes their spectroscopy very efficient. As a result, MOS is a critical tool for extracting the potentially transformational scientific yield from the RC surveys currently undergoing a huge increase in sensitivity on the pathway to the Square Kilometre Array.

\acknowledgements DJBS would like to thank the organisers for their invitation to give a talk, and to acknowledge the contributions of the WEAVE and WEAVE-LOFAR science teams. DJBS wishes to particularly thank Ian Heywood for making reduced J-VLA data products available prior to publication, as well as Matt Jarvis, Huub R\"ottgering and Chris Simpson for useful discussions. DJBS also acknowledges a financial award from the Santander Universities Partnership Scheme. 

\bibliography{DSmith} 

\end{document}